\documentclass[twocolumn,prl]{revtex4}

\let\textcite\relax
\usepackage{fancyhdr}
\makeatletter
\DeclareRobustCommand{\MakeUppercase}[1]{{%
      \def\i{I}\def\j{J}%
      \def\reserved@a##1##2{\let##1##2\reserved@a}%
      \expandafter\reserved@a\@uclclist\reserved@b{\reserved@b\@gobble}%
      \protected@edef\reserved@a{\uppercase{#1}}%
      \reserved@a
   }}
\DeclareRobustCommand{\MakeLowercase}[1]{{%
      \def\reserved@a##1##2{\let##2##1\reserved@a}%
      \expandafter\reserved@a\@uclclist\reserved@b{\reserved@b\@gobble}%
      \protected@edef\reserved@a{\lowercase{#1}}%
      \reserved@a
   }}
\makeatother
\expandafter\let\csname ver@natbib.sty\endcsname\relax

\usepackage{amsmath,amssymb,amsfonts,amsthm}

\usepackage{enumerate}
\usepackage{enumitem}
\usepackage{epsfig}
\usepackage{latexsym}
\usepackage{xcolor}
\usepackage[colorinlistoftodos, shadow, textsize=small, obeyDraft]{todonotes}

\definecolor{lightblue}{rgb}{0.68, 0.85, 0.9}
\definecolor{airforceblue}{rgb}{0.36, 0.54, 0.66}

\usepackage[colorlinks=true,citecolor=blue,linkcolor=blue,urlcolor=airforceblue]{hyperref}
\usepackage{comment}

\usepackage{todonotes}

\usepackage[backend=bibtex8,url=true,eprint=true,doi=true,sorting=none,backref=true]{biblatex}
\newcommand{\bite}{\begin{itemize}}
\newcommand{\eat}{\end{itemize}}
\newcommand{\beq}{\begin{equation}}
\newcommand{\eeq}{\end{equation}}

\newcommand{\beqa}{\begin{align}}
\newcommand{\eeqa}{\end{align}}
\newcommand{\beqar}{\begin{eqnarray}}
\newcommand{\eeqar}{\end{eqnarray}}
\newcommand{\barr}{\begin{array}}
\newcommand{\earr}{\end{array}}

\newcommand{\expect}[1]{\langle #1\rangle}

\newcommand{\fullket}[1]{\vert #1 \rangle}

\newcommand{\ie}{\emph{i.e. }}

\begin{document}

\preprint{This line only printed with preprint option}

\title{Thermal Time and Kepler's Second Law\footnote{This article is dedicated to my wife, my constant companion in all my endeavors.}}

\author{Deepak Vaid}
\email{dvaid79@gmail.com}
\affiliation{National Institute of Technology, Karnataka}

\date{\today}

\begin{abstract}
It is shown that a recent result regarding the average rate of evolution of a dynamical system at equilibrium in combination with the quantization of geometric areas coming from LQG, implies the validity of Kepler's Second Law of planetary motion.
\end{abstract}

\maketitle

\section{Introduction}\label{eqn:intro}
One of the leading contenders for a theory of quantum gravity is the field known as Loop Quantum Gravity (LQG) \cite{Vaid2014LQG-for-the-Bewildered}. A fundamental result of this approach towards reconciling geometry and quantum mechanics, is the quantization of geometric degrees of freedom such as areas and volumes \cite{Rovelli1993Area,Rovelli1994Discreteness}. LQG predicts that the area of any surface is quantized in units of the Planck length squared $ l_p^2 $.

Despite its many impressive successes, in for instance, solving the riddle of black hole entropy \cite{Krasnov1996Counting,Rovelli1996Black} or in understanding the evolution of geometry near classical singularities such as at the Big Bang or at the center of black holes \cite{Ashtekar2011Loop,Banerjee2012Introduction,Bojowald1999Loop}, LQG is yet to satisfy the primary criteria for any successful theory of quantum gravity - \emph{agreement between predictions of the quantum gravity theory and well known and well understood aspects of spacetime as described by classical physics}. The same is also true of the other, more popular, approaches to quantum gravity such as String Theory and its offshoot the AdS/CFT correspondence. This correspondence does offer predictions about the behavior of the quark-gluon plasma generated in heavy-ion collisions \cite{Natsuume2014AdS/CFT,Policastro2001Shear,Chirco2010The-universal,Kovtun2005Viscosity,Policastro2002From,Son2002Minkowskispace,Son2007Viscosity}, however, as with black hole entropy or the big bang these predictions concern energy scales far outside the reach of our daily experience.

\begin{figure}[tbph]
\centering
\includegraphics[width=0.7\linewidth]{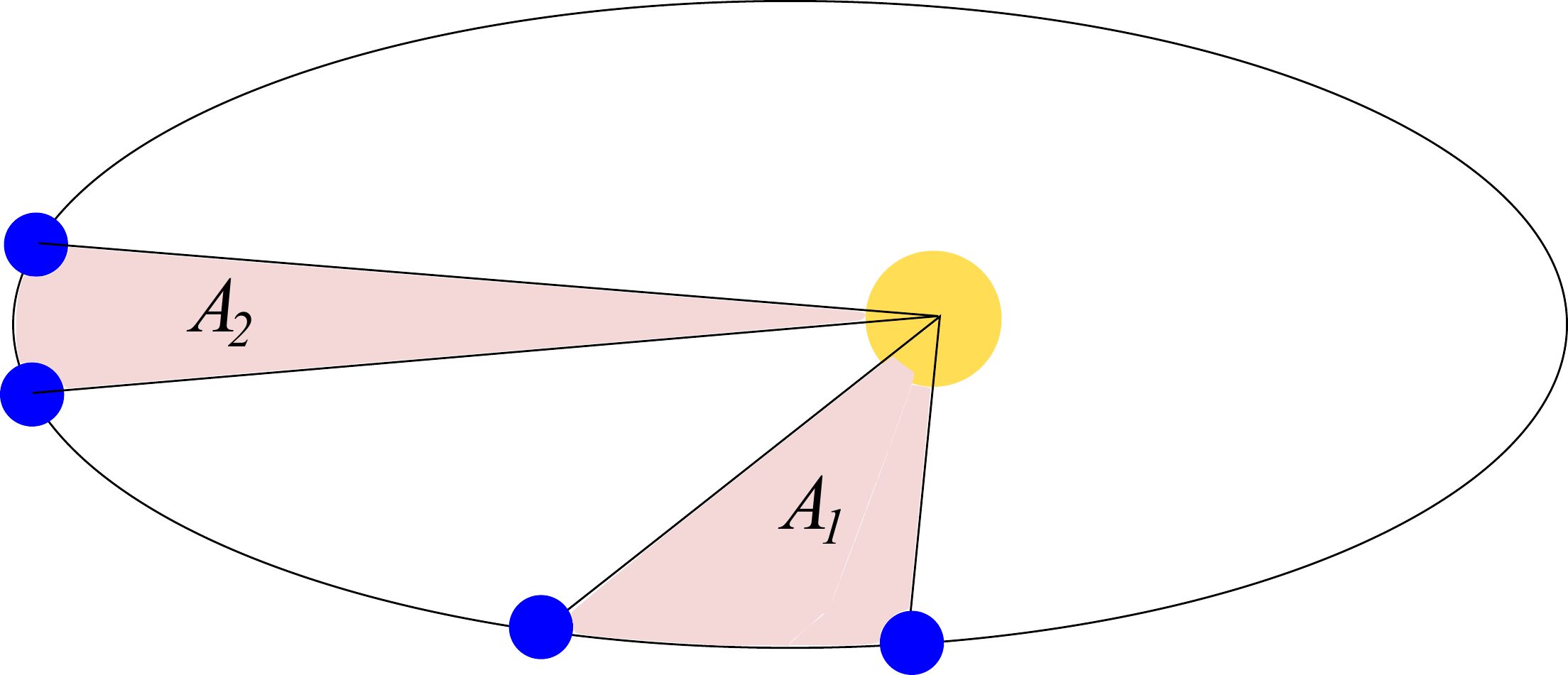}
\caption{Kepler's Second Law of Planetary Motion: $ A_1, A_2 $ are the areas swept out by the line joining the planet to the Sun. If the two segments are traversed in equal time interval then $ A_1 = A_2 $}
\label{fig:equalareas}
\end{figure}

Here we argue that a recent result \cite{Haggard2013Death} regarding the \emph{average} rate at which a quantum system at equilibrium transitions between (nearly) orthogonal states, when taken together with the quantization of geometric areas arising from LQG, implies the validity of Kepler's Second Law of planetary motion (\autoref{fig:equalareas}). This would be a prediction from a theory of quantum gravity which has a \emph{direct} correspondence with an \emph{observable} and well-understood aspect of classical gravity.
\section{Maximum Speed of Quantum Evolution}\label{eqn:maxspeed}
In \cite{Haggard2013Death}, Haggard and Rovelli (HR) discuss the relationship between the concept of \emph{thermal time}, the Tolman-Ehrenfest effect and the rate of dynamical evolution of a system - \ie the number of distinguishable (orthogonal) states a given system transitions through in each unit of time. The last of these is also the subject of the Margolus-Levitin theorem \cite{Margolus1998Maximum} according to which the rate of dynamical evolution of a macroscopic system with fixed average energy (E), has an \emph{upper bound} ($\nu_{\perp}$) given by:
\begin{equation}
\label{eqn:margolus-levitin}
\nu_{\perp} \leq \frac{2E}{h}
\end{equation}

Note that $\nu_{\perp}$ is the maximum possible rate of dynamical evolution, not the \emph{average} or \emph{mean }rate, and also that in \cite{Margolus1998Maximum} the bound is determined by the \emph{average energy}: $E = \sum_n c_n E_n \fullket{\psi_n}$ of a system in a state $\fullket{\Psi} = \sum_n c_n \fullket{\psi_n}$, where $\{\fullket{\psi_n}\}$ is a basis of energy eigenstates of the given system. As a corollary the \emph{smallest} time interval which the system can be used to measure, if used as a clock, is given by
\begin{equation}\label{eqn:ml-timestep}
	t^{ML}_{\perp} = h/2E
\end{equation}

What HR work with is not the average energy of the system, but the fluctuation around the mean given by:
\begin{equation}\label{eqn:e-fluctuation}
\Delta E^2 = \expect{ H^2 } - \expect{ H } ^2
\end{equation}
where $H$ is the Hamiltonian of the given system. Thus the elementary time-step that HR consider is given by:
\begin{equation}\label{eqn:hr-timestep}
t^{HR}_{\perp} = h/2\Delta E
\end{equation}
There is a big difference between the two time-steps $t^{HR}_{\perp}$ and $t^{ML}_{\perp}$. The former depends on the standard deviation in the mean energy of a given system, while the latter depends on the mean energy itself. The difference can be seen via the following argument given on pg. 8 of ML's paper, and I quote.
\begin{quote}
	For an isolated, macroscopic system with energy average energy $E$, one can construct a state which evolves at a rate $v=2E/h$. If we have several non-interacting macroscopic subsystems, each with average energy $E_i$, then the average energy of the combined system is $E_{tot} = \sum_i E_i$ ... our construction applies perfectly well to this combination of non-interacting subsystems for which we can construct a state which evolves at a rate $v_{\perp} = 2E_{tot}/h = \sum_i 2E_i/h = \sum_i v^i_{\perp}$. Thus if we subdivide our total energy between separate subsystems, the maximum number of orthogonal states per unit time for the combined system is just the sum of the maximum number for each subsystem taken separately.
\end{quote}
Therefore for a composite system, consisting of a number of non-interacting subsystems each with maximum rate of evolution $\nu^{ML}_i$, the maximum rate of evolution is simply the sum of the rates of the individual subsystems: $\nu^{ML} = \sum_i \nu^{ML}_i$. As the system size increases the minimum time-interval and the maximum rate of dynamical evolution associated with the system also increase. the Margolus-Levitin bound is, thus, an \emph{extensive} property of a system.

Now, while there is no restriction on the average energy of each of the subsystem - we can pick a subsystem that is as small or as large as we wish - the same is not true for the variance of the energy of each subsystem. In fact, if each of the subsystems is in equilibrium with all the other subsystems and with the composite system as a whole, then the variance of the energy for each subsystem must be the same and also be equal to that of the composite system.

Thus, whereas, Margolus-Levitin tell us that increasing the energy of a system increases the (maximum) rate of dynamical evolution, the rate given by Haggard-Rovelli depends only on the temperature $kT \sim \Delta E$, a quantity which does not depend on system size (in a suitable thermodynamic limit). Of course, Haggard-Rovelli do take due care to state that their elementary time-step is the \emph{average} time the system takes to move from one state to the next.

\section{Kepler's Second Law}\label{sec:keplerslaw}
Having gone over the distinction between the Margolus-Levitin theorem and the Haggard-Rovelli result, let us move to consider how one can possibly apply Haggard and Rovelli's reasoning to a real world problem, namely that of two-body central force problem in the Newtonian theory of gravitation. There we have \href{https://en.wikipedia.org/wiki/Kepler%27s_laws_of_planetary_motion}{Kepler's three laws} deduced empirically, which preceded Newton's analytic solution of the two-body problem. The three laws are:
\begin{enumerate}
 	\item The orbit of every planet is an ellipse with the Sun at one of the two foci.
 	\item A line joining a planet and the Sun sweeps out equal areas during equal intervals of time.
 	\item The square of the orbital period of a planet is directly proportional to the cube of the semi-major axis of its orbit.
\end{enumerate}

For the time being, let us assume that the two-body Sun+Planet system is a thermal system which happens to be at equilibrium and whose dynamical evolution is given by the motion of the planet around the Sun (or more precisely, the motion of the planet and the sun around \emph{each other}). Now, according to HR the number of states swept out by such a system in the course of its evolution, over a given interval of time $\delta t$ is the same regardless of the location of the system along its dynamic trajectory. This is a quantity which depends only on the characteristic temperature $kT$ of the system.

\begin{figure}[tbph]
\centering
\includegraphics[width=0.7\linewidth]{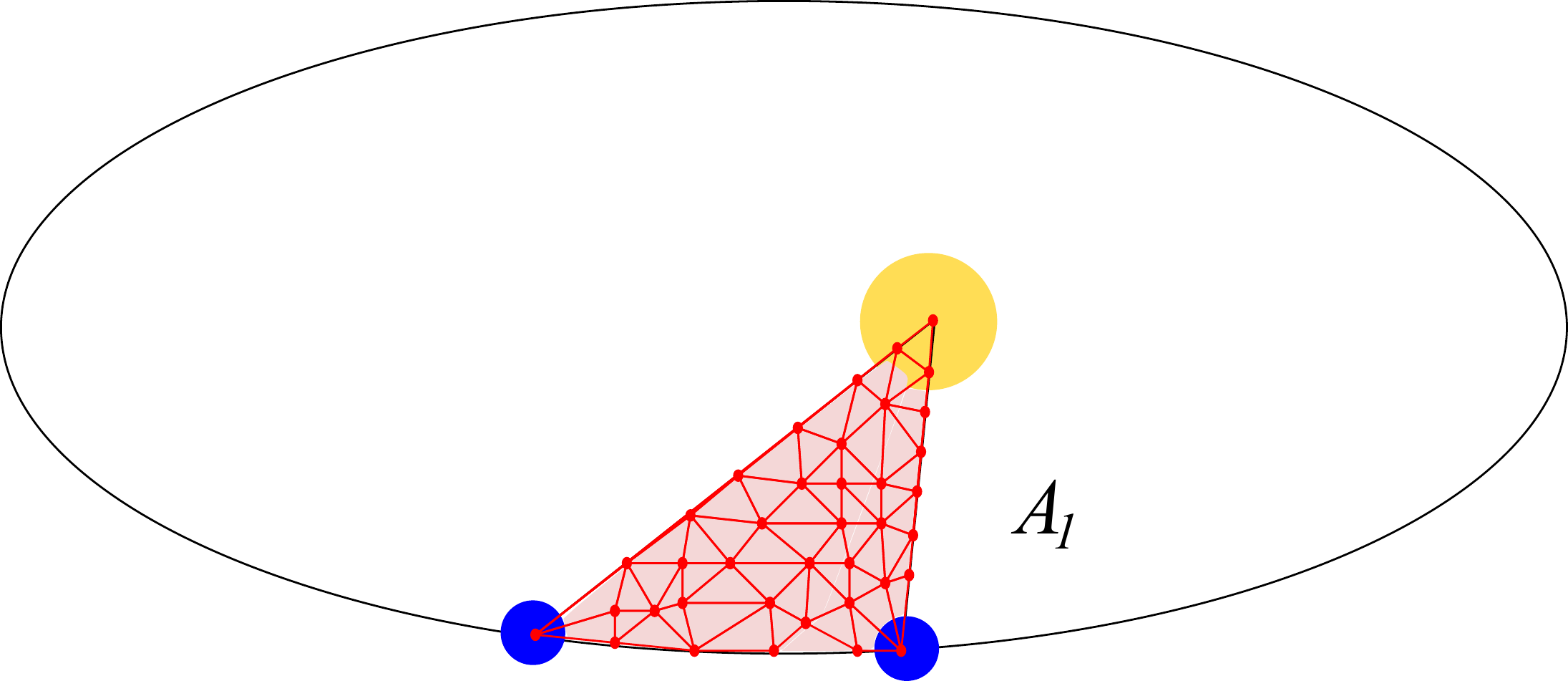}
\caption{LQG tells us that the macroscopic area $ A_1 $ can be decomposed into smaller quanta. Each triangle in the triangulation of $ A_1 $ corresponds to a single quantum of area.}
\label{fig:quantizedarea}
\end{figure}

For the two-body system, the ``states'' in question correspond to the various microscopic configurations of quantum geometry which the system traverses over a given time interval $\delta t$. The macroscopic area value $\delta A$ swept out by the planet's trajectory during $\delta t$ is a sum over all the microscopic quantum geometric area elements which make up the patch under consideration (\autoref{fig:quantizedarea}). $\delta A$ is thus a measure of the total number of microscopic states the system passes through during $\delta t$. Consequently, Haggard-Rovelli tell us that the number of such states the system passes through per unit time is a constant:
\begin{equation}\label{eqn:kepler-two}
\frac{\delta A}{\delta t} \propto kT
\end{equation}
\emph{which is nothing other than Kepler's second law of planetary motion}. Thus Haggard and Rovelli's result, along with our understanding of the nature of the microscopic degrees of freedom of quantum geometry allows us to derive one of the most basic results of Newtonian physics - Kepler's second law - and relate it to the thermodynamics of the many-body system consisting of the quantum geometric degrees of freedom which determine the macroscopic dynamics of the system.

\section{Conclusion}

The notion of gravity as an emergent or induced force is certainly not new and can be traced back some fifty years ago to Sakharov's original paper \cite{Sakharov1968Vacuum} on this topic. Since then the emergent or induced gravity paradigm has resurfaced in many different forms: gravity arising from defects in a world crystal \cite{Kleinert1990Gauge,Kleinert2005Emerging}; gravity arising from the equilibrium thermodynamics of apparent horizons seen by accelerated observers \cite{Jacobson1994Black,Padmanabhan2002Thermodynamics,Padmanabhan2002The-Holography,Padmanabhan2003Gravity,Padmanabhan2012Equipartition,Padmanabhan2012Emergence,Padmanabhan2012Emergent,Padmanabhan2015Gravity-and}; gravity as the effective low-energy dynamics of various condensed matter systems \cite{Laughlin2003Emergent,Hu2005Can-Spacetime,Samuel2006Surface,Volovik2008Emergent,Sindoni2009Emergent,Liberati2009Analogue,Xu2010Emergent,Hamma2010A-quantum,Caravelli2011Trapped,Dreyer2012Internal,Vaid2013Non-abelian,Vaid2013Superconducting}; gravity arising from the entanglement entropy of an underlying microscopic many body system \cite{Van-Raamsdonk2010Building,Jacobson2012Gravitation,Swingle2012Constructing,Matsueda2013Emergent,Raasakka2016Spacetime-free,Cao2016Space,Jacobson2016Entanglement}; and finally gravity as an entropic force \cite{Padmanabhan2009Equipartition,Verlinde2010On-the-Origin,Padmanabhan2015Distribution,Padmanabhan2015Gravity-and}.

The present work also falls within the emergent gravity paradigm. Here the assertion is that the properties of classically gravitating systems can be understood by applying treating macroscopic geometry as a many body system composed of quanta of geometry in thermal equilibrium.

Several gaps in the argument have yet to be filled in. Most important of these is the assertion that a system of two heavenly bodies orbiting each other can be viewed as an equilibrium configuration of an underlying many-body quantum gravitational system. There is no reason why this should \emph{not} be the case. Black hole geometries are by now well-understood to correspond to such equilibrium states. It would only be natural that as our understanding of quantum gravity deepens, geometries more general than those of black holes will be understood as equilibrium (or near equilibrium) states.

It is possible that the present argument when applied to the question of the rotation curves of stars in galactic disks might shed light on the anomalous behavior of these curves which, at present, requires the postulate of dark matter for its resolution. This, however, remains in the realm of speculation and the subject of a possible line of future research.

(Note: The arguments in this article were first presented in a \href{http://www.physicistatwork.com/blog/2013/02/11/thermal-time-and-keplers-second-law/#more-43}{blog post} by the author in early 2013.)

\printbibliography
\end{document}